\documentclass[a4paper,fleqn]{cas-dc}

\usepackage[authoryear,longnamesfirst]{natbib}
\usepackage[utf8]{inputenc}
\usepackage[T1]{fontenc}

\begin{document}


\shorttitle{}    


\title[mode = title]{Simulation of combined radio and radar signals at the Radar Echo Telescope for Cosmic Rays}  



%

\author[1]{K. Nivedita}[orcid=0000-0002-5301-3177]
\fnmark[1]
\ead{krishna.gopinath@ru.nl}

\author[2]{I. Loudon}
\author[3]{J. Loonen}
\author[4]{P. Allison}
\author[4]{J.J. Beatty}
\author[5]{D.Z. Besson}
\author[4]{A. Connolly}
\author[6,7,8]{A. Cummings}
\author[9]{C. Deaconu}
\author[3]{S. de Kockere}
\author[3]{K.D. de Vries}
\author[10]{I. Esteban}
\author[3]{D. Frikken}
\author[11]{C. Hast}
\author[12]{E. {Huesca Santiago}}
\author[13]{C.-{Y. Kuo}}
\author[2]{A. Kyriacou}
\author[3]{U.A. Latif}
\author[3]{V. Lukic}
\author[5]{C. McLennan}
\author[1,14]{K. Mulrey}
\author[13]{J. Nam}
\author[5]{S. Prohira}
\author[5]{J.P. Ralston}
\author[5]{M.F.H. Seikh}
\author[1]{N. Shahid}
\author[3]{R.S. Stanley}
\author[3]{J. Stoffels}
\author[2]{S. Toscano}
\author[3]{D. {Van den Broeck}}
\author[3]{N. {van Eijndhoven}}
\author[6,4]{S. Wissel}



\affiliation[1]{Department of Astrophysics,IMAPP, Radboud University, P.O. Box 9010, 6500 GL Nijmegen, The Netherlands}
\affiliation[inst2]{
    organization={Université libre de Bruxelles (ULB), Faculté des Sciences},
    addressline={CP 230},
    postcode={1050},
    city={Brussels},
    country={Belgium}
}
\affiliation[3]{Vrije Universiteit Brussel, HEP@VUB, IIHE, Brussels, Belgium}


\affiliation[4]{Department of Physics, Center for Cosmology and AstroParticle Physics, The Ohio State University, Columbus, OH 43210, USA}

\affiliation[5]{Department of Physics and Astronomy, University of Kansas, Lawrence, KS 66045, USA}

\affiliation[6]{Center for Multi-Messenger Astrophysics, Institute for Gravitation and the Cosmos, Pennsylvania State University, University Park, PA 16802, USA}

\affiliation[7]{Department of Physics, Pennsylvania State University, University Park, PA 16802, USA}

\affiliation[8]{Department of Astronomy and Astrophysics, Pennsylvania State University, University Park, PA 16802, USA}

\affiliation[9]{Department of Physics, Enrico Fermi Institute, Kavli Institute for Cosmological Physics, University of Chicago, Chicago, IL 60637, USA}

\affiliation[10]{
    organization={Department of Physics and EHU Quantum Center, University of the Basque Country UPV/EHU},
    addressline={P.O. Box 644},
    city={Bilbao},
    postcode={48080},
    country={Spain}
}

\affiliation[11]{SLAC National Accelerator Laboratory, Menlo Park, CA 94025, USA}

\affiliation[12]{Deutsches Elektronen-Synchrotron DESY, Platanenallee 6, 15738 Zeuthen, Germany}

\affiliation[13]{Department of Physics, Graduate Institute of Astrophysics, Leung Center for Cosmology and Particle Astrophysics, National Taiwan University, Taipei, Taiwan}
\affiliation[14]{Nikhef, Science Park Amsterdam, 1098 XG Amsterdam, The Netherlands}


\cortext[1]{Corresponding author}

\begin{abstract}
To explore neutrino astronomy at high energies (> 10 PeV), the Radar Echo Telescope for Cosmic Rays (RET-CR) was developed to assess the feasibility of a radar technique for detecting particle cascades in ice, serving as a precursor to the Radar Echo Telescope for Neutrinos (RET-N). The main concept of RET-CR is that, as a high-energy cosmic-ray air-shower core propagates into the high-altitude ice sheet, a dense secondary-particle cascade is created, which is very similar to that of an in-ice high-energy neutrino-induced cascade. At RET-CR, the expected signal consists of three distinct components: radio emission from the in-air particle shower, Askaryan radio emission from the secondary in-ice cascade, and the radar signal itself arising from the reflection of the transmitted radio signal from the ionisation trail of the in-ice secondary cascade. In this work, we present the first combined simulation-based package and study aimed at characterising the combined radio and radar signals at the in-ice receivers at RET-CR. We describe the simulation framework and provide a detailed discussion of the salient features of the radio and radar signals, including their spatial footprints and temporal characteristics, as predicted for the shallow in-ice detectors.
\end{abstract}

\maketitle

\section{Introduction}
Neutrinos are unique cosmic messengers with exceptional potential to probe the extreme universe, as they can travel vast cosmological distances unaffected by magnetic fields or intervening matter. The IceCube Neutrino Observatory achieved the first detection of astrophysical neutrinos in the TeV–PeV energy range in 2013, marking a major milestone in the development of neutrino astronomy \citep{article1,article2,article3,article4}. More recently, the KM3NeT collaboration reported the detection of a nearly horizontal, high-energy cosmic neutrino event (KM3-230213A), which currently represents the highest-energy neutrino observed to date, with a corresponding muon detected at an estimated energy of 120 PeV \citep{article5}.

A broad range of experimental efforts and novel detection techniques are underway to extend the sensitivity to even higher energies, reaching into the EeV regime, and to the cosmic neutrino flux \citep{Ackermann2022}. These initiatives are predominantly driven by radio-based neutrino detection methods, which exploit the coherent radio emission produced by neutrino-induced particle cascades and offer a scalable approach to instrumenting the large target volumes required at ultra-high energies \citep{article9,article10}.  An example of an in-ice radio-based neutrino detection experiment, the Radio Neutrino Observatory (RNO-G) is located on top of the Greenland ice sheet \citep{article11}.
\subsection{The Radar Echo Telescope}
The Radar Echo Telescope (RET) explores a new radar-based detection technique for high- and ultra-high-energy neutrinos. The primary objective is to detect the ionisation trail left behind by a high-energy neutrino-induced particle cascade in ice using radar reflections.  This technique has been successfully tested at the laboratory in the SLAC-T576 experiment \citep{article12,article14}. To explore the scope of this method in nature, a pathfinder experiment, the Radar Echo Telescope for Cosmic Rays (RET-CR), was deployed at the Greenland Summit Station during the summers of 2023 and 2024. At RET-CR, high-energy cosmic-ray air showers that propagate into the high-altitude ice sheet are used as a test beam for radar reflections. \citep{article13,article15}.

Summit Station is at an altitude of 3200 m, so the top of the icesheet is closer to the shower maximum  ($X_{\text{max}}$) , and thus, a large fraction of the shower can propagate into the ice. Thus, detecting this secondary cascade from air showers in ice with an in-ice radar system could verify the radar method and pave the way for advancement towards the Radar Echo Telescope for Neutrinos (RET-N).
\begin{figure}
  \centering
  \includegraphics[width=.96\columnwidth]{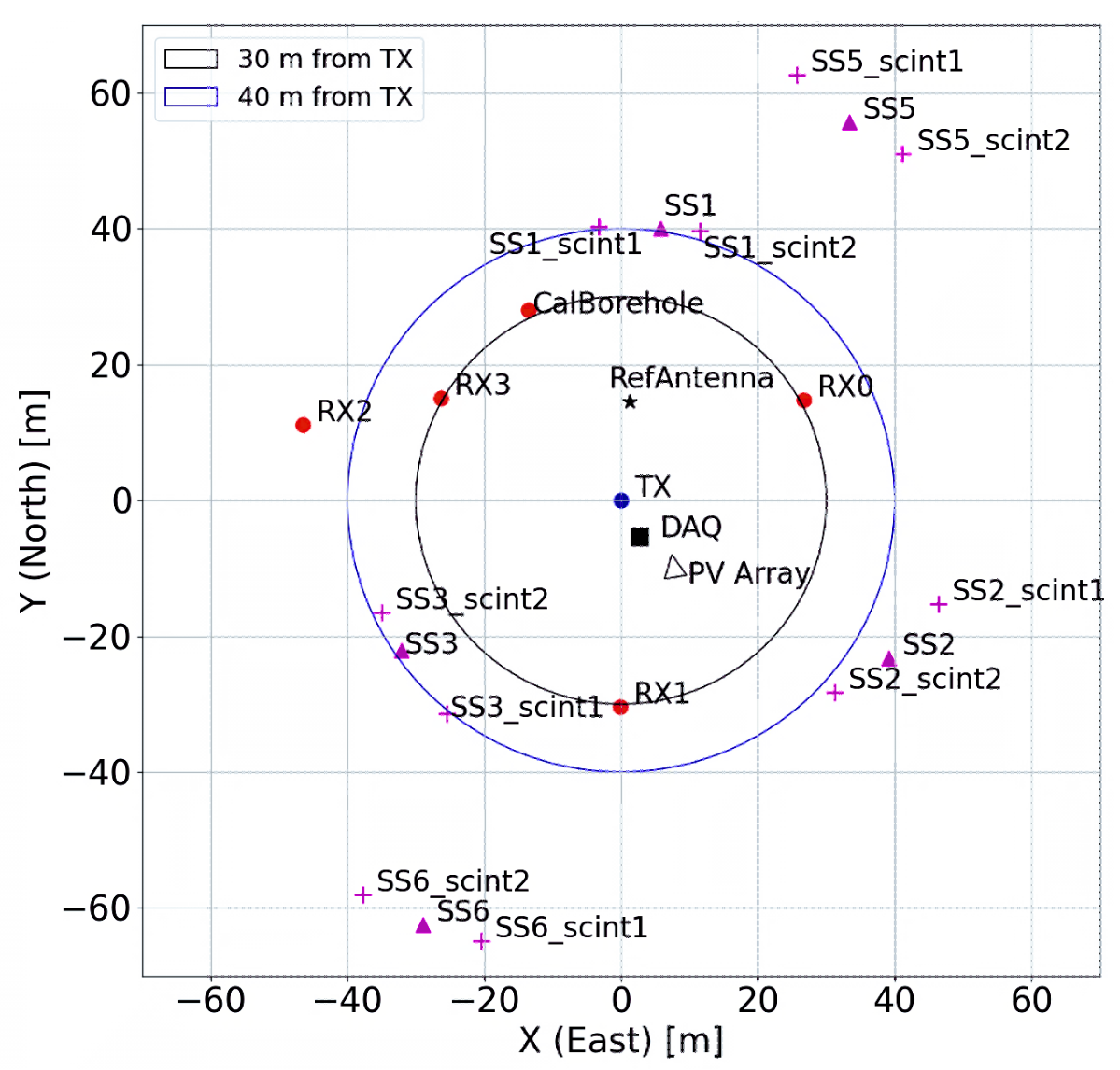}
  \caption{The RET-CR experimental layout (taken from GNSS satellite survey) is depicted. The surface stations (SS) (($+$) denotes the scintillator panels, while the solid triangles ($\blacktriangle$) denote the radio antennas), and the in-ice radar system (TX and RX) are labelled.}
  \label{fig:exp_lay}
\end{figure}

RET-CR completed a full data-taking run in summer 2024, recording on the O($10^5$) cosmic-ray triggered events \citep{Frikken2024}. The experimental layout is shown in Figure \ref{fig:exp_lay}, and comprises of five surface stations and an in-ice radar-echo system. The surface stations of RET-CR consist of Icetop scintillator panels and LPDAs (Log Periodic Dipole Array) of 50 - 350 MHz \citep{article29}. The radar system comprises a phased array with eight transmitter antennas, with a maximum output of of 20 W each and four receiver antennas. The surface stations trigger on incoming cosmic-ray air showers, providing a trigger for the in-ice component, and use their own independent reconstruction strategy \citep{article27}. Meanwhile, the radar system aims to detect the air-shower-induced secondary cascade in ice associated with the same cosmic-ray event.  
\\
\\
Extensive air showers initiated by high-energy cosmic rays emit coherent radio signals in the 10-100's of MHz range as they propagate through the atmosphere. The in-air radio emission results from two mechanisms: the geomagnetic effect \citep{article16,Scholten2009}, caused by the deflection of charged particles in Earth’s magnetic field, and the Askaryan (charge-excess) effect, due to a net negative charge buildup in the shower front \citep{article6}.  When the particles of an air shower,  such as the high-energy electrons, positrons, and photons, propagate into the ice, they produce a compact particle cascade characterised by a net negative charge excess. This net charge leads to coherent radio emission, the Askaryan emission, that is strongly forward-beamed around the Cherenkov angle and spans frequencies from hundreds of MHz to the GHz range \citep{article7,deVries2016,DeKockere2022,ara}.

For a single cosmic ray event, in addition to the reflected radar signal from the secondary cascade, the receivers are expected to detect radio emission from the particle cascades themselves. These include the in-air radio emission produced by the extensive air shower, as well as the in-ice Askaryan emission generated by the secondary cascade in ice. 

In this work, we present our first combined simulation package for the RET-CR experiment, incorporating both in-air and in-ice radio emissions along with radar signals. The presence of complementary radio signatures provides valuable information for interpreting the RET-CR data and cross-validating the detected signals. This article is organised as follows: we first present the simulated broadband radio and radar electric field signals for a set of representative shower geometries to illustrate the signal properties. We then investigate a realistic treatment with a simple detector response, bandpass filtering, and thermal noise. This demonstrates that restricting the bandwidth around the transmitter frequency suppresses much of the broadband radio emission compared to the narrowband radar echo. Finally, we conclude towards implications of these results in the search for radar echoes in RET-CR data.

\section{Simulation Framework}

The radio emissions from both in-air and in-ice particle cascades are simulated using FAERIE \citep{article17}, which incorporates CORSIKA along with CoREAS \citep{article18}, and Geant4 \citep{article19}. The radar signals are simulated using MARES \citep{article20}. Also, note that throughout this study we use the CORSIKA-based coordinate system, as described in \citep{article21}.

\subsection{Radio emission from particle cascades }

FAERIE is a software framework that propagates particle-level information from a cosmic ray air shower simulated with CORSIKA  into a Geant4-based ice medium. Particles within a radius of 100 cm around the shower core are passed into the ice to model the development of the in-ice secondary cascade and its associated Askaryan radio emission. We use an ice profile modelled from measurements taken near the Summit Station \citep{article23} in Geant4.  The CORSIKA air showers are generated using site-specific parameters, including a realistic geomagnetic field configuration and altitude (see the Appendix \ref{appendix1} for more details).

FAERIE uses the endpoint formalism \citep{article22,Zilles2017}, also employed in CoREAS, to simulate radio signals as they would appear at the antennas beneath the ice surface. Within FAERIE, a new parameter, \textit{IceBoundaryAltitude}, is introduced to define the altitude of the air-to-ice boundary; in this work, it is set to 3.2 km, corresponding to the elevation of the RET-CR site.   

\subsection{Radar simulations}
MARES is a semi-analytical C++ framework that models radar echoes from particle cascades using a macroscopic description of the cascade ionisation profile, coupled with radar scattering equations.

The RET-CR in-ice transmitter system employs a phased array of eight antenna elements. During the 2024 data-taking run, each element was nominally operated with an output power of $5~\mathrm{W}$ and a forward antenna gain of $2.3~\mathrm{dB}$, corresponding to an effective power of approximately $8.5~\mathrm{W}$ per element. Since coherent phasing causes the transmitted power to scale as $\text{N}^2$, where N is the number of phased elements, the total transmitted power is approximately $136~\mathrm{W}$ for four elements and $544~\mathrm{W}$ for all eight elements, when they are phased coherently. These two configurations are used in the MARES simulations. The simulations are run at 182.16 MHz, corresponding to the transmitter frequency during the RET-CR data taking run. For the simulations, we also used a plasma lifetime of 10 ns  \citep{article20} .

\subsection{Timing correlations}
To compare the radio and radar signals from the same cosmic-ray event, a common timing reference must be established. In CORSIKA, particle times are given relative to the moment the primary cosmic ray enters the atmosphere, whereas CoREAS defines $t=0$ as the time when the shower front reaches the observer's altitude, which, for in-ice antennas, is the ice surface. 

To ensure consistency, the particle times from CORSIKA are shifted to the CoREAS convention before generating the secondary cascade in Geant4. The radar signal timings from MARES are referenced to the start of the secondary cascade as $t=0$, allowing a direct comparison between the radio and radar signals.

Depending on the shower geometry, the radio and radar signals may arrive either well separated in time or partially overlap. Any measurable delay between the two could help distinguish the radar echo from the in-air and in-ice radio emission, providing additional information about the cosmic-ray event.

\subsection{Energy deposition in ice - MARES and FAERIE}
\label{sec:energy_dep}

\begin{figure}
    \centering
    \includegraphics[width=.999\columnwidth]{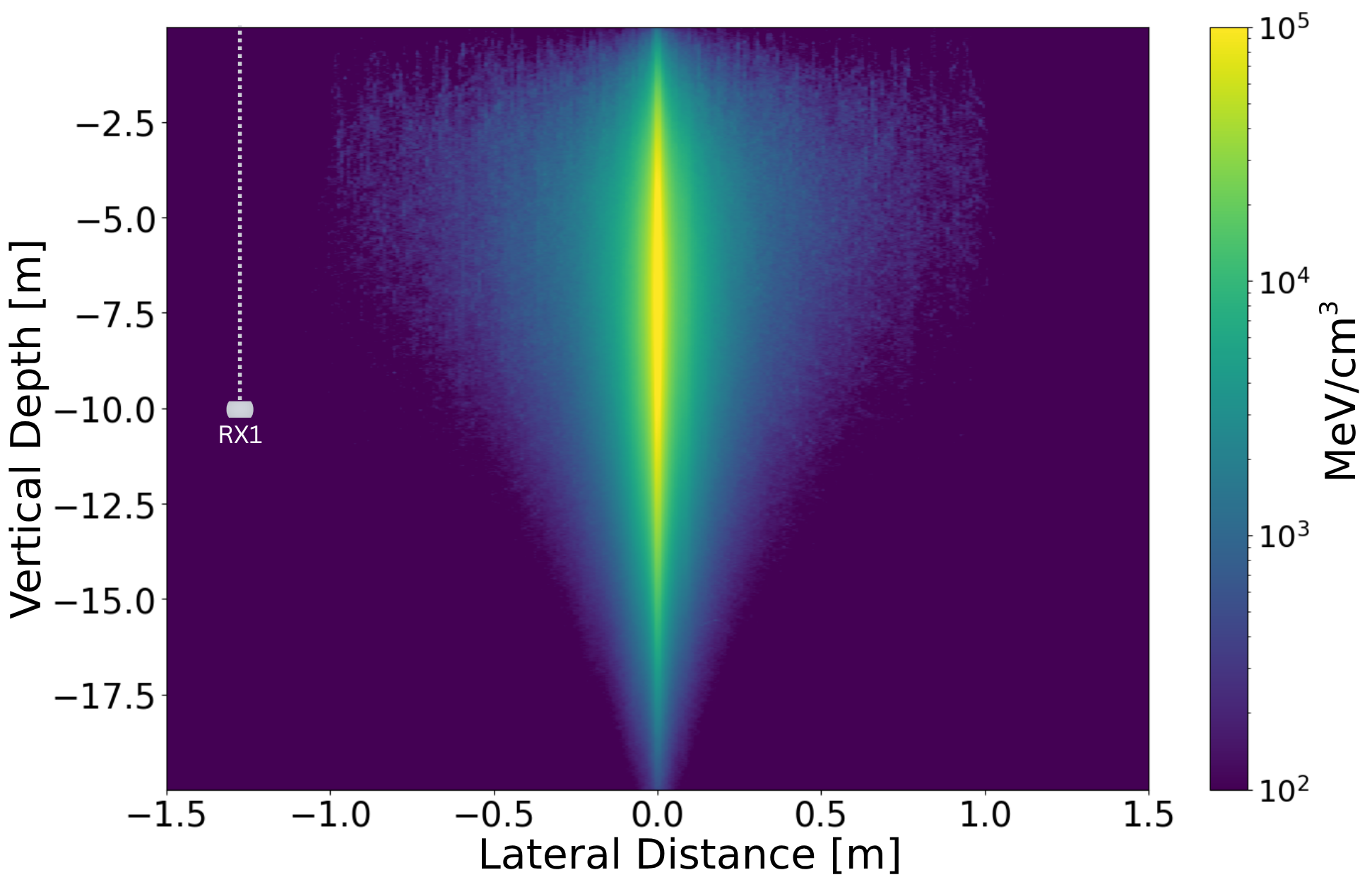}
    \caption{Simulated energy density profile for an in-ice secondary cascade from Geant4,  from a cosmic--ray proton primary of $10^{17}~$eV. A receiver at close proximity (within 1.5 m) to the shower vertex is also displayed to demonstrate RET-CR receiver antenna depths with the in-ice cascade geometry.}
    \label{fig:energydep}
\end{figure}

Cosmic-ray air showers of PeV to EeV energies generated by Monte Carlo simulations will exhibit shower-to-shower fluctuations, whereas MARES uses an analytical method based on the Nishimura-Kamata-Greisen (NKG) parametrisation \citep{article26,article30,loudon_modelling_2026} to generate target cascades for a particular primary energy. 
Therefore, for a single CR event, it is necessary to ensure that the parameters of the air shower and the resulting in-ice cascade produced by FAERIE are propagated into MARES, so that the three signal components are comparable and can be accurately combined. 
In order to achieve this, simulations are run in the order described below: 
\begin{itemize}
    \item First, each event is initiated with CORSIKA, where the event parameters - primary energy, particle type, arrival direction - are defined, and the air shower is created. 
    \item The particle output from CORSIKA is then fed into FAERIE/Geant4 to propagate the shower into the ice volume, generating both the resulting in-ice radio emission and the in-ice cascade profile.
    \item Finally, the energy deposit profile from Geant4 is used to inform the parameters of the analytic cascade generated by MARES, producing the radar reflection. 
\end{itemize}
This method allows us to ensure that, to a close approximation, the radar signal and the two radio components are produced by the same CR air shower.  An illustration of the energy deposit in the ice by the cascade is depicted along a vertical slice in Figure \ref{fig:energydep}.
\subsection{Approximations and caveats}
In the shallow geometry considered in this work, the observer may be  located at distances comparable to the longitudinal extent of the particle cascade, with receiver depths only a few meters below that of shower maximum. Under these conditions, the far-field approximation, with coherent Askaryan collimated into a few degree-wide Cherenkov cone, is not generally applicable, resulting in significantly greater radio power at off-cone angles than prescribed in standard parameterisations such as AMVZ \citep{Jaime2006coherent}. Instead, radiation emitted from different sections of the cascade reaches the observer with different path lengths and phases, producing a geometry-dependent interference pattern\citep{article28}.

Additionally, the current implementation of MARES considers either the direct or the surface-reflected propagation path for a given simulation. Simultaneous treatment of multiple propagation paths is not yet implemented. Consequently, the coherent interference between direct and reflected electromagnetic fields prior to scattering from the particle cascade is not modelled.
\section{Combined signal study for RET-CR}
\begin{figure*}[!htbp]
    \centering
    \includegraphics[width=\linewidth]{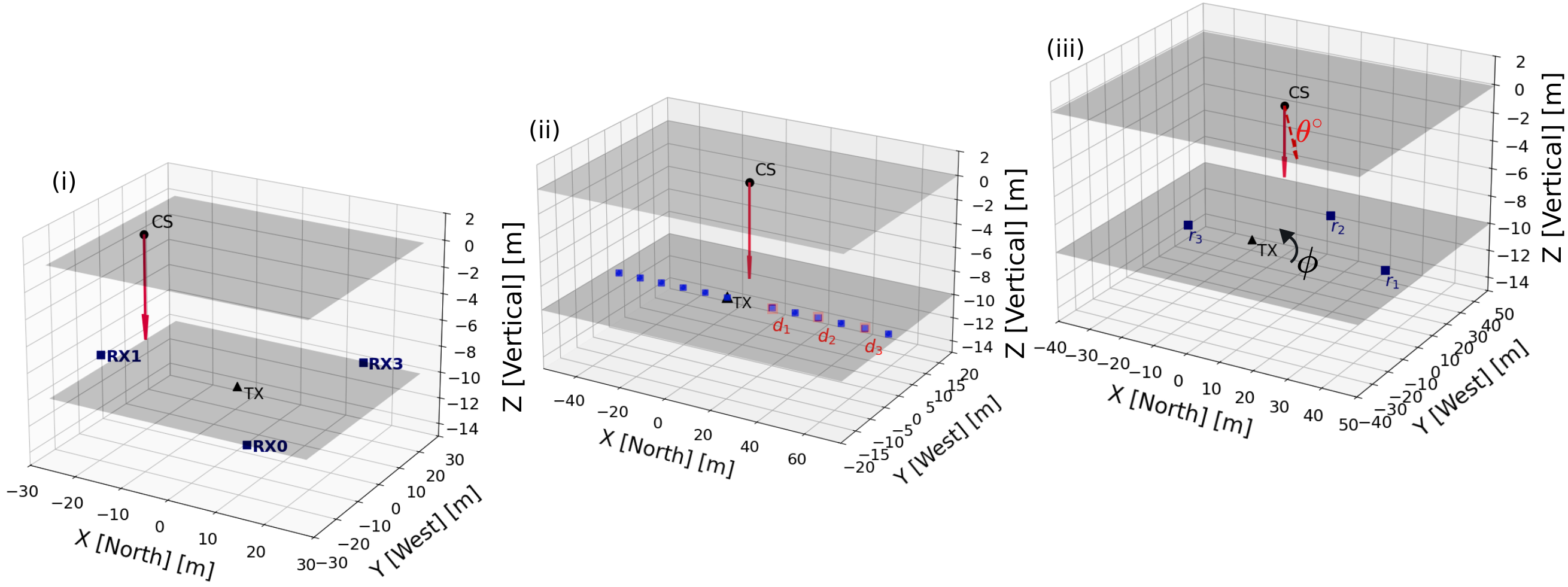}
    \caption{
Simulation configurations considered in this work.
\textbf{(i)} Representative RET-CR geometry for a vertical $10^{17}$~eV proton-induced air shower with shower core located at $(-20,0,0)$~m. The radar system follows the RET-CR layout shown in Figure~\ref{fig:exp_lay}. \textbf{(ii)} Vertical $10^{17}$~eV proton-induced air shower with shower core at $(10,0,0)$~m. The observer locations are marked by grey squares at
$X = -60$ to $60$~m in steps of $10$~m. The selected observer locations are at $d_1 = 10$~m, $d_2 = 30$~m, and $d_3 = 50$~m, for which the waveforms will be presented.
\textbf{(iii)} $10^{18}$~eV proton-induced air shower with shower core at $(10,0,0)$~m and observer locations $r_1$, $r_2$, and $r_3$, used for the arrival-direction study. In all the geometries, the transmitter is located at [0,0,-10].
}
\label{fig:configuration}
\end{figure*}
To study the combined radio and radar signals, we consider three representative simulation geometries as shown in Figure  \ref{fig:configuration}. We first examine a representative RET-CR event and compare the characteristics of the radio and radar signals across all polarisations. We then present a more detailed analysis of the signal arrival times and arrival-direction dependence, focusing on the vertical polarisation, to which the in-ice receivers are most sensitive. The results presented in the first two cases correspond to cosmic-ray air showers with a primary energy of $10^{17}$~eV, whereas the final geometry assumes a primary energy of $10^{18}$~eV.

\subsection{A representative RET-CR event}
\begin{figure*}[!htbp]
    \centering
    \includegraphics[width=0.9\textwidth]{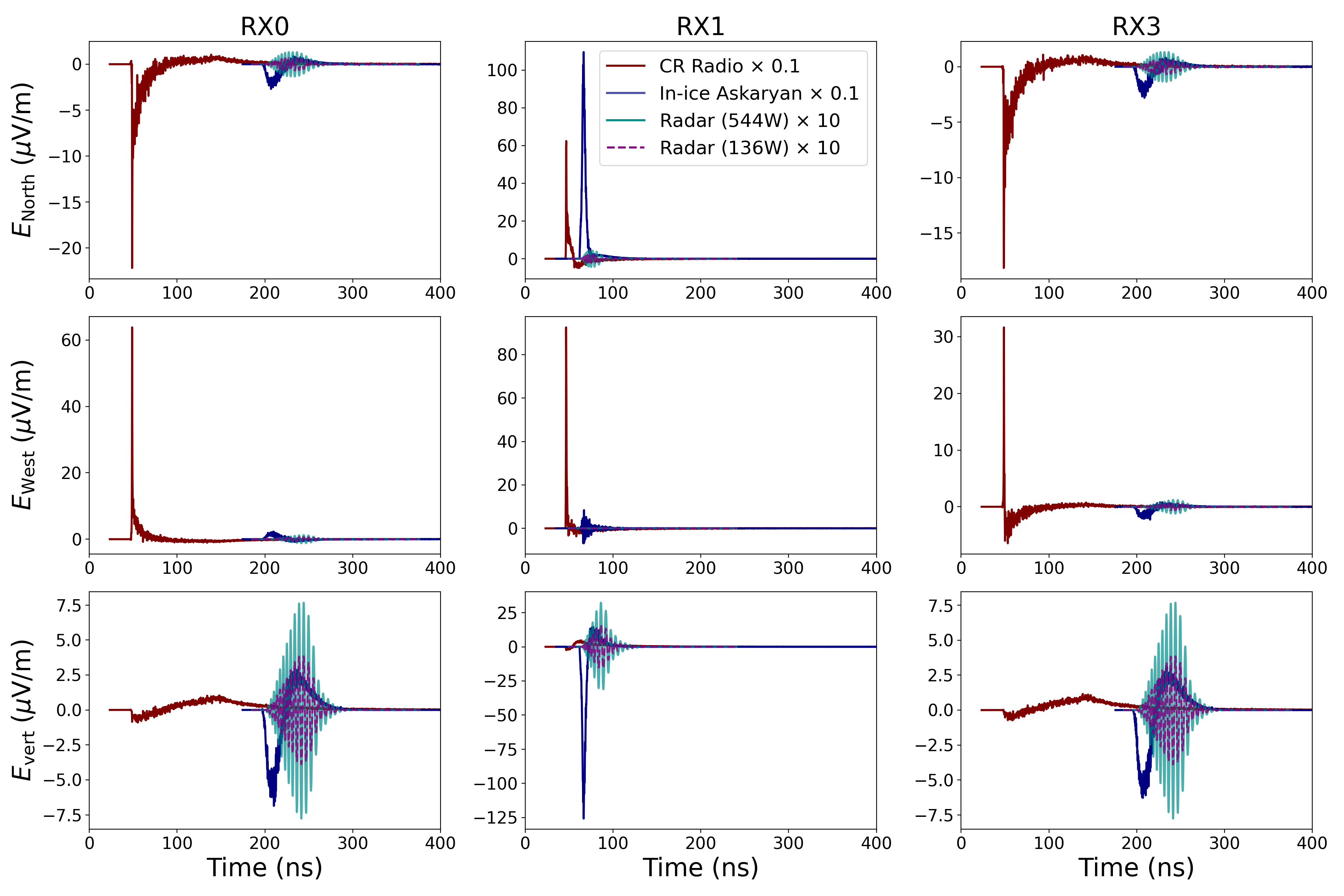}
    \caption{Combined radio and radar signals simulated for all polarizations at the three RET-CR receivers for the geometry shown in Figure \ref{fig:configuration}(i).  The amplitudes are scaled for clarity of presentation and are not directly comparable in this figure.}
    \label{fig:case_study}
\end{figure*}

\begin{figure*}[!htbp]  
    \centering
    \includegraphics[width=0.9\textwidth]{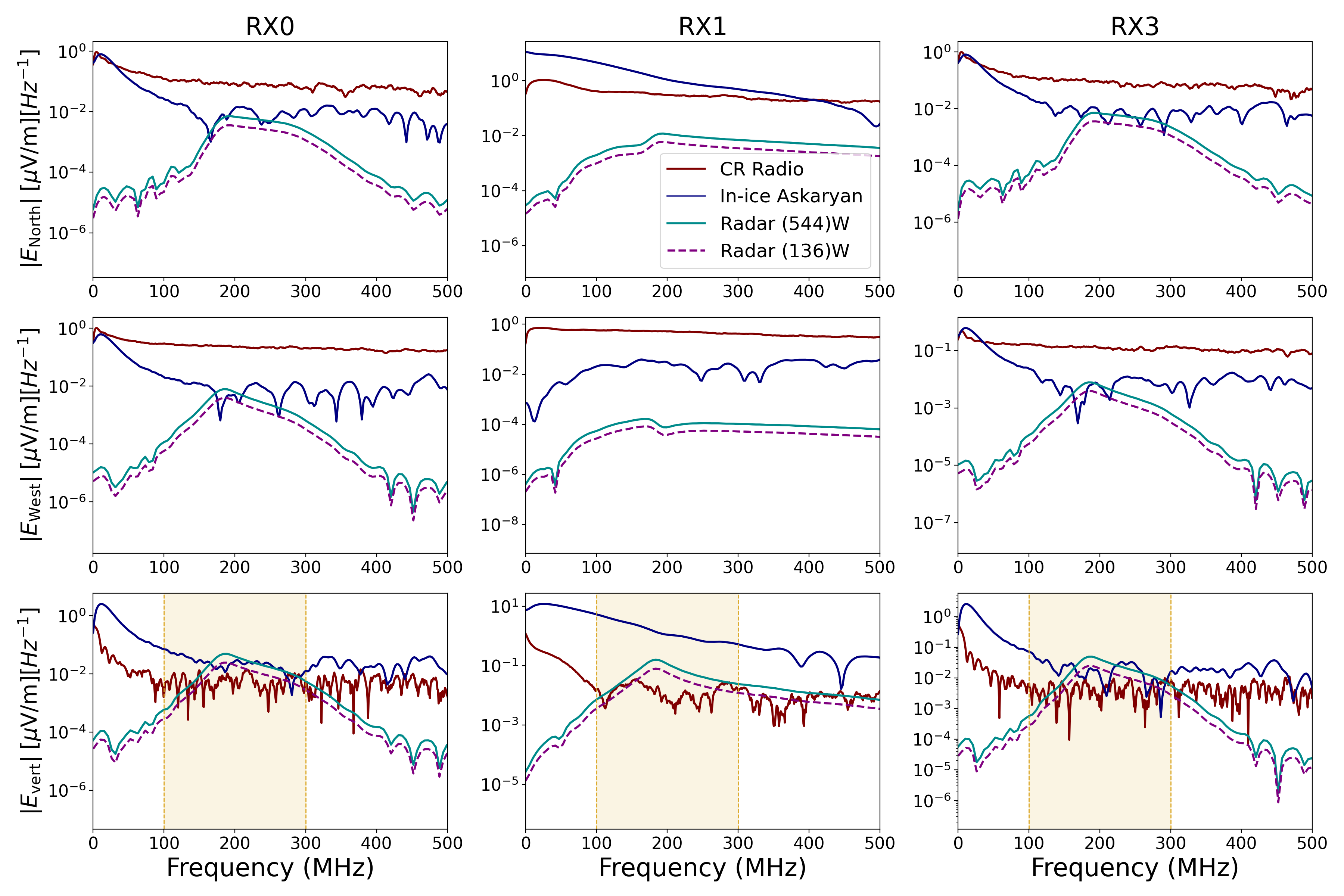}
    \caption{Combined radio and radar spectrum simulated for all polarizations at the three RET-CR receivers for the geometry shown in Figure \ref{fig:configuration}. The corresponding electric fields are in Figure \ref{fig:case_study}. RET-CR receiver bandwidth is
indicated by the yellow band.}
    \label{fig:case_study_spec}
\end{figure*}

 In the representative geometry shown in Figure \ref{fig:configuration}(i), the shower vertex of the secondary in-ice cascade is located at $(-20.0~\mathrm{m},0.0~\mathrm{m})$ in the North-West coordinate system at the ice surface. The three selected RET-CR receivers,  RX0 $(14.8~\mathrm{m},-26.7~\mathrm{m})$, RX1 $(-30.4~\mathrm{m},0.1~\mathrm{m})$, and RX3 $(15.1~\mathrm{m},26.3~\mathrm{m})$, are placed in the simulations at a depth of $10~\mathrm{m}$, following the RET-CR array configuration. The simulated event corresponds to a vertical cosmic-ray air shower with a primary energy of $100~\mathrm{PeV}$.  Figure \ref{fig:case_study} shows the combined electric-field traces from the in-air radio emission, the in-ice Askaryan emission, and the radar echo. The corresponding spectra are shown in Figure \ref{fig:case_study_spec}.  For visual clarity, the in-air and in-ice Askaryan signals are scaled by a factor of 0.1, while the radar signals are scaled by a factor of 10. In this geometry, all the signals reach RX1 first, then RX0 and RX3, reflecting the relative positions of the receivers with respect to the emission region. 
\\
\\
For the in-air cosmic-ray radio emission, the electric field is dominated by the East–West component, as expected from geomagnetic radiation. The in-air emission always arrives at the receivers earlier than the in-ice Askaryan and radar components. The cosmic-ray radio emission originates from higher up in the atmosphere and is thus observed in the (very) forward direction, leading to strong coherence up to high frequencies, as shown in Figure \ref{fig:case_study_spec}, and also the vertical component of the in-air emission is significantly weaker than the horizontal components. 
\\
\\
The in-ice Askaryan emission is radially polarised. This is observed in RX1, which lies approximately along the North-South direction relative to the cascade vertex, and records strong North-South and vertical radio components and retains coherence to higher frequencies in the spectrum. In contrast, RX0 and RX3 lie on opposite sides of the cascade in the East-West direction, resulting in opposite polarities in the East-West radio component. 

Unlike in-air radio emission, Askaryan emission in ice shows a strong vertical radio component. Since the RET-CR receivers are located at shallow depths, they are closer in depth to the shower maximum of the secondary in-ice cascade. Due to this, the geometrical effects associated with the cascade's longitudinal extent can become more relevant than for detector configurations that are deeper in-ice. Although radiation emitted along the shower axis can still add coherently near the Cherenkov angle, the observer receives contributions from different parts of the cascade at varying distances, leading to a geometry-dependent interference pattern. Consequently, a well-defined Cherenkov cone for the in-ice Askaryan signal may not always be observed at these shallow observer distances \citep{article28}. 

From a simple geometrical estimate for this configuration, considering the Cherenkov angle in ice to be, $\theta_c = 45^\circ$ near the firn, and a shower maximum at a depth of $5~\mathrm{m}$, gives $\theta_c-\theta_v \simeq 19^\circ$ for RX1 and $\theta_c-\theta_v \simeq 40^\circ$ for RX0 and RX3, where $\theta_v$ is the viewing angle between the line from the shower maxima to the receivers and the shower axis. Since RX1 is located closer to the Cherenkov angle, its signal is expected to retain stronger coherence, and RX0 and RX3 will have weaker coherence towards higher frequencies. 
\\
\\
The radar echoes show a different behaviour. Since the simulated transmitter emits vertically polarised fields at $182.16~\mathrm{MHz}$, the reflected radar signals are strongest in the vertical component.  The signal properties of the radar echoes largely depend on their location within the transmitter-cascade-receiver geometry.

For this event, the radar echo arrives close in time to the Askaryan signal in ice. This motivates a careful study of the signal and spectral features in order to distinguish between the different emission components, especially when multiple radio-emission mechanisms contribute simultaneously. As observed in Figure \ref{fig:case_study_spec}, we have a stronger radar emission (in the vertical polarisation), over the in-ice Askaryan close to the RET transmitter frequency (182 MHz), for observers at RX0 and RX3, particularly at the transmitted power of 544 W for this geometry. 

\subsection{Signal properties and air shower geometry}

\begin{figure*}[!htbp]  
    \centering
    \includegraphics[width=\textwidth]{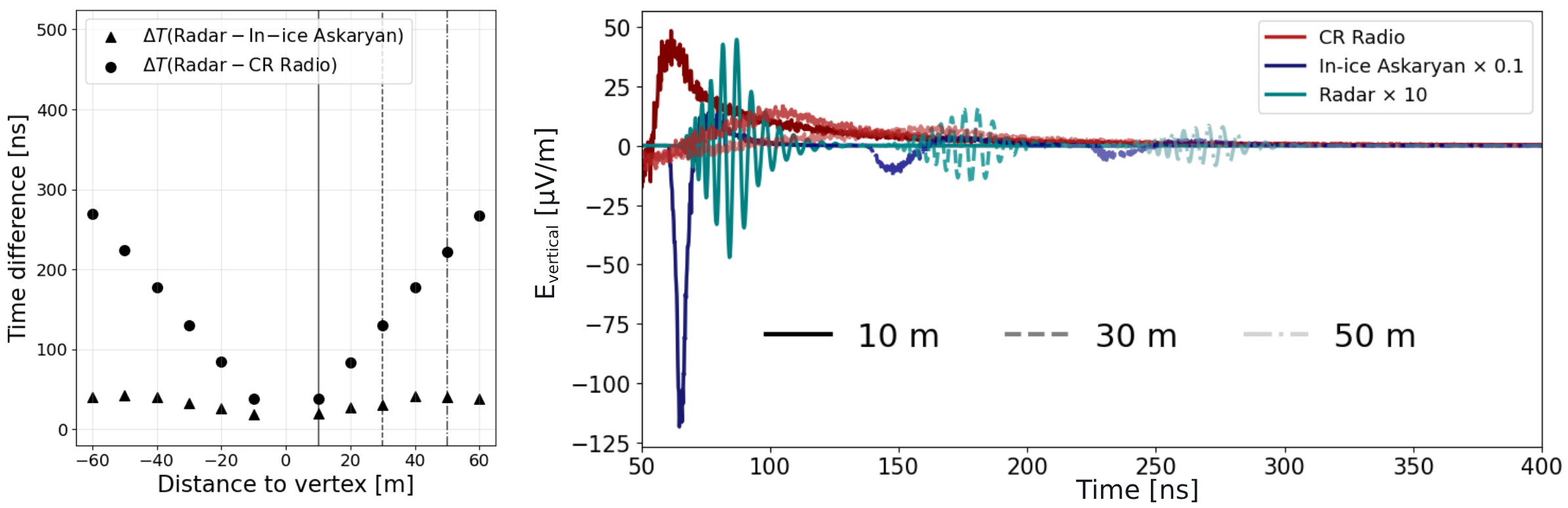}
    \caption{Left: Arrival-time difference as a function of observer position relative to the shower vertex, evaluated for various observer locations between $-60$ m and $+60$ m. The dashed vertical lines indicate the three representative observer positions at 10 m, 30 m, and 50 m from the vertex. Right: Simulated signals for these three observer locations. The corresponding geometry is shown in Figure~\ref{fig:configuration}(ii). The amplitudes are scaled for clarity of presentation and are not directly comparable in this figure.}
    \label{fig:time_vertex}
\end{figure*}
The timing and spectral content of the signals are strongly dependent on the air shower geometry. The relative timing between the radar echo and the radio emissions provides an additional handle to constrain key shower parameters, such as the arrival direction, and the location of the in-ice cascade vertex or the position of the air-shower core. For timing and spectral studies, we examine the vertical polarisation of the electric fields, as our in-ice receiver dipole antennas are most sensitive to the vertically polarised signals.

\begin{figure*}[!htbp]
    \centering
    \includegraphics[width=\textwidth]{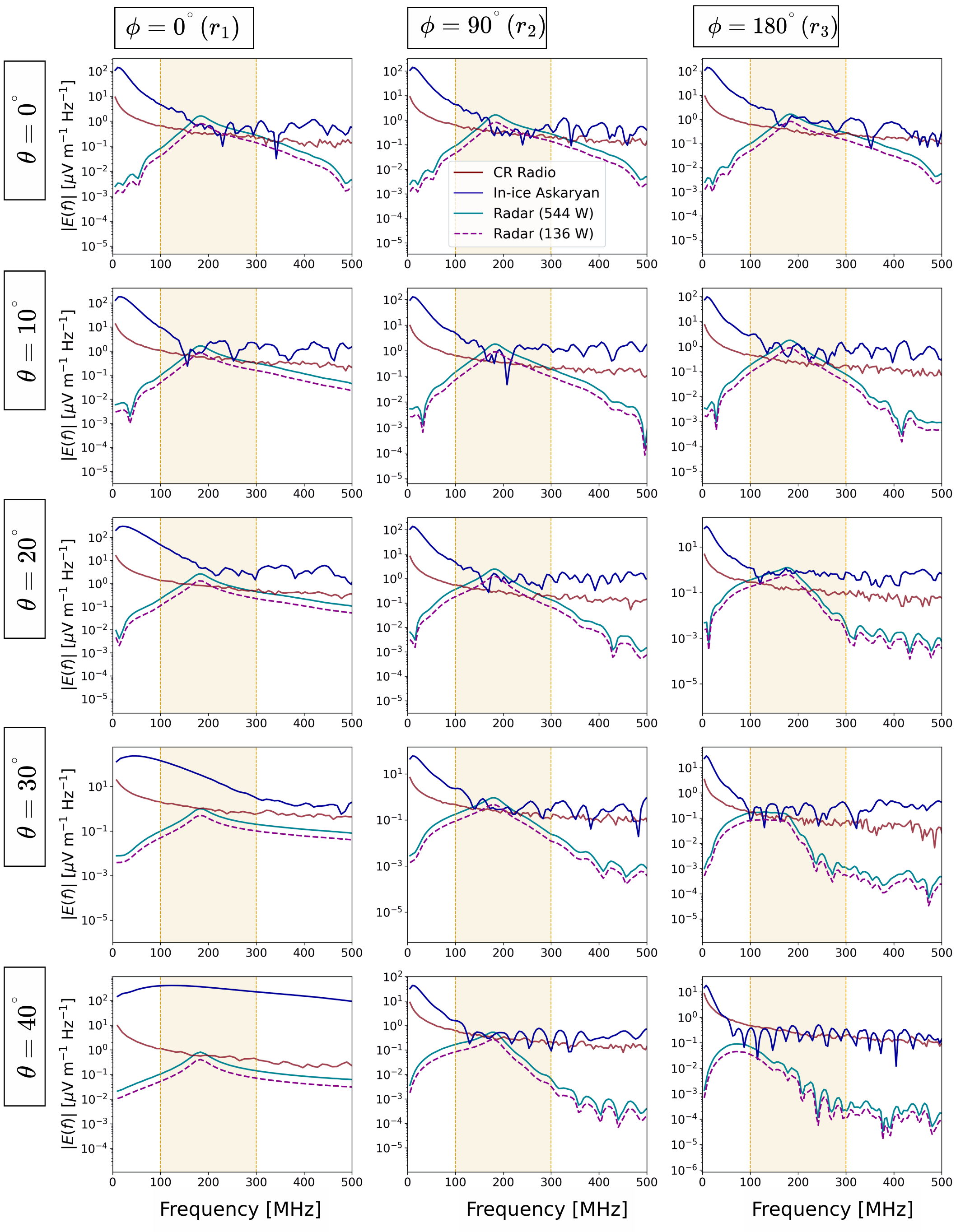}
   \caption{Simulated vertically polarised electric-field spectra at receiver locations corresponding to azimuthal angles $\phi = 0^\circ$, $90^\circ$, and $180^\circ$ relative to the in-ice cascade, for air showers with zenith angles $\theta = 0^\circ$--$40^\circ$ (top to bottom). The detector geometry is shown in Figure~\ref{fig:configuration}(iii). The receivers are located on the ice surface, 30~m from the shower vertex at $[10,0,0]$, with the transmitter at $[0,0,-10]$. The primary energy is $10^{18}$~eV, transmitter powers are 136~W and 544~W, and the RET-CR receiver bandwidth is indicated by the yellow band.}
    \label{fig:spec_all}
\end{figure*}
\subsubsection{Arrival times}
Figure \ref{fig:time_vertex} shows the relative arrival-time differences between the cosmic-ray radio emission and the in-ice Askaryan signal, relative to the radar echoes, as a function of distance from the shower vertex. We simulate a vertical shower with a primary energy of 100 PeV, and observers at varying distances from the shower vertex. Figure \ref{fig:time_vertex} also presents the corresponding electric field signals from three receiver antennas at $d_1,d_2$ and $d_3$ at distances of 10 m, 30 m and 50 m (along the +X direction) from the shower vertex. The arrival time is defined as the time at which the peak electric-field amplitude of the dominant signal polarisation occurs.

The cosmic-ray in-air radio signal consistently arrives before both the in-ice Askaryan signal and the radar echo. As the receiver distance from the shower vertex increases, the time separation between the cosmic-ray in-air radio signal and the radar echo also increases. The radar and Askaryan signals remain temporally close, with arrival-time differences typically ranging from approximately 10 to 60 ns, depending on the receiver position. The time delay between the cosmic-ray in-air radio emission and the radar signal provides a useful observable for constraining the shower vertex position and narrowing the search window for radar echoes.

The in-ice radio and radar echo signals are simulated using identical ray-tracing methods, yielding similar arrival times. Since the arrival time is defined by the peak of each signal, differences in pulse shapes can also introduce additional offsets in this estimate. 

\subsubsection{Arrival directions}
The arrival direction of the primary cascade impacts the frequency content of the received radar signals. A detailed characterisation of the total signal as a function of arrival direction is therefore conducted to enable the separation of the overlapping emission components.

To understand the combined signals, we consider the geometry setup in Figure \ref{fig:configuration}(iii). 
We simulated a cosmic ray primary of $10^{18}$ eV, considering a range of zenith angles ($\theta$) from $0^\circ$ to $40^\circ$ at an azimuthal direction of $\phi=0^\circ$ (pointing to the North). We show the simulated signal spectra for all configurations in Figure \ref{fig:spec_all}. The simulated time-domain signals are shown in Appendix \ref{appendix3} in Figure \ref{fig:time_all}. 

In the setup in Figure \ref{fig:configuration}(iii), three receiver locations (labelled as \textit{$r_1$}, \textit{$r_2$} , \textit{$r_3$}), are illustrated. The first location $r_1$ (40, 0) is positioned in the forward direction of an inclined secondary cascade (along $\phi = 0^\circ$) , the second location $r_2$ (10, 30) is placed perpendicular to the cascade axis ($\phi = 90^\circ$) (along the +Y, or, the West direction) , while the third location $r_3$ (-20, 0)  lies in the backward facing direction the cascade inclination ($\phi = 180^\circ$) (along the -X, or, the South direction).  All these locations are positioned 30 m laterally around the shower vertex [X, Y, Z] = [10, 0, 0], and all the antennas are at a depth of 10 m in ice. 
\begin{figure*}[!htbp]
    \centering
    \includegraphics[width=\linewidth]{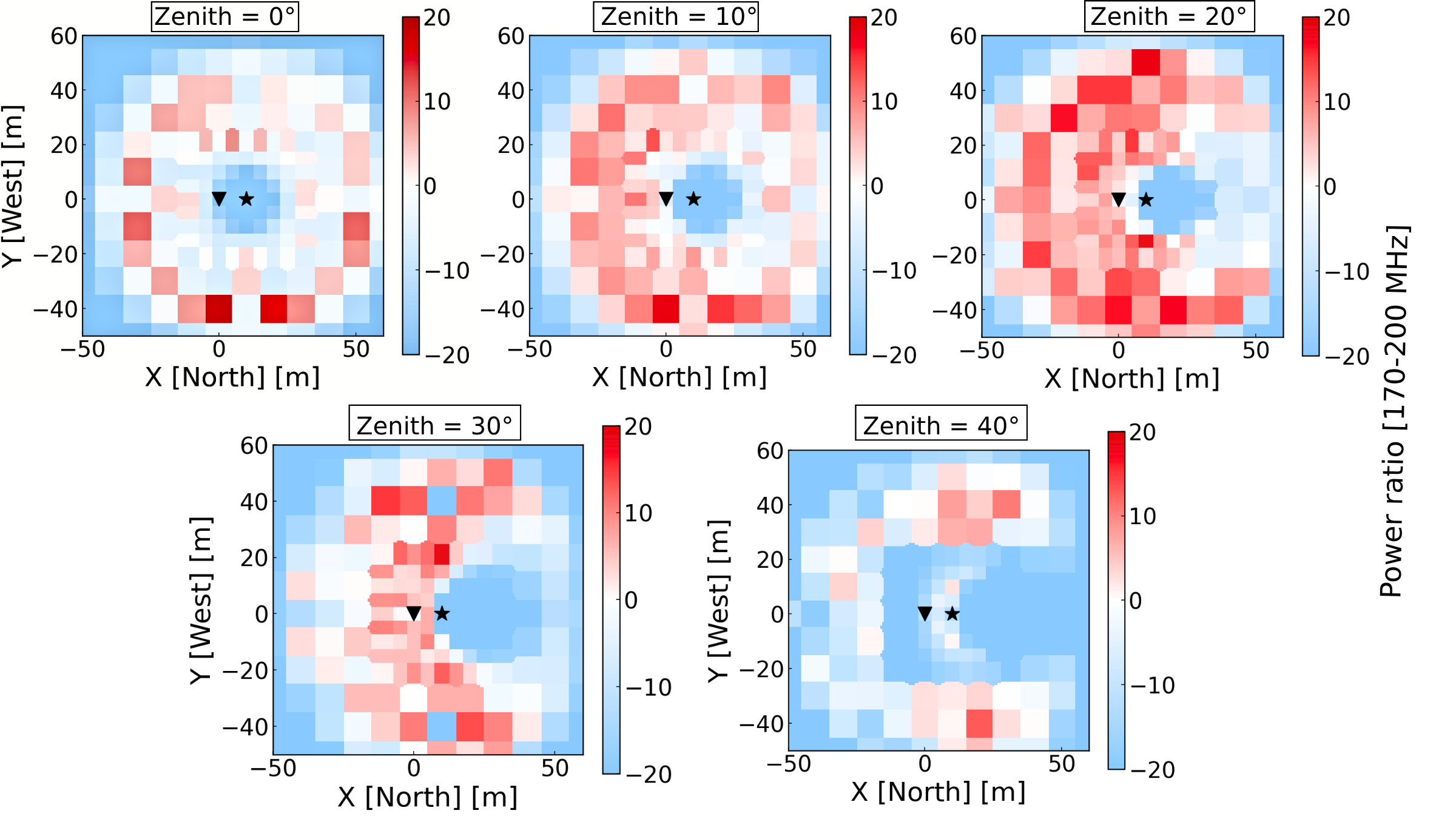}
    \caption{Heat map that depicts regions with higher radar emission in a narrow bandwidth of 170-200 MHz. Zenith angles for the secondary cascade are $0^\circ$,   $10^\circ$,  $20^\circ$, $30^\circ$, and $40^\circ$. The transmitter (TX) is indicated by ($\blacktriangle$), while the shower core is marked by ($\bigstar$). }
    \label{geoms}
\end{figure*}

For inclined cascades ($\theta > 10^\circ$), the observer location at $r_1$ lies closer to the coherent emission region of the in-ice Askaryan signal. Consequently, the in-ice radio emission retains coherence over a broad frequency range, extending to higher frequencies and becoming dominant over the radar echo contribution. This enhanced coherence is particularly evident for zenith angles between $20^\circ$ and $40^\circ$, where the Cherenkov cone intersects the observer locations (see Figure~\ref{fig:fluence} in Appendix \ref{appendix3}), resulting in a substantial increase in the high-frequency radio signal. 

Meanwhile, for a vertical cascade ($\theta = 0^\circ$) and a slightly inclined cascade ($\theta = 10^\circ$), an antenna located at $r_1$ remains sufficiently away from the Cherenkov cone at a distance of 30$~$m. In these geometries, the radar echo becomes visible within the RET bandwidth in the forward-facing direction at 544W power. 

At the second location, $r_2$,  the radar echoes are visible over both the in-air and in-ice radio emission around the transmitted frequency,  for all simulated zenith angles. 

Finally, for the third location, $r_3$,  the radar signals show reduced amplitudes due to the Doppler shift towards lower frequencies. For higher inclinations above $30^\circ$, the radar echo amplitudes will be further reduced below the in-ice Askaryan amplitudes within the RET bandwidth.

Figure \ref{fig:spectrographs_noradar} and \ref{fig:spectrographs_radar} are presented in the Appendix \ref{appendix3}, and show the spectrograms in the 140–220 MHz frequency range for two signal cases: the radio-only signal for comparison and the combined signal including the radar contribution. A frequency window with a width of 80 MHz centred on the transmitter frequency of 182.16 MHz is plotted here to highlight features associated with the radar echoes. A dashed line is also plotted at the transmitter frequency. In the spectrogram, for certain geometries, particularly at larger zenith angles at $r_1$, an additional early-time signal is visible between 50--100 ns, which corresponds to the in-air radio emission component. 
\begin{figure*}[!htbp]
    \centering
    \includegraphics[width=\textwidth]{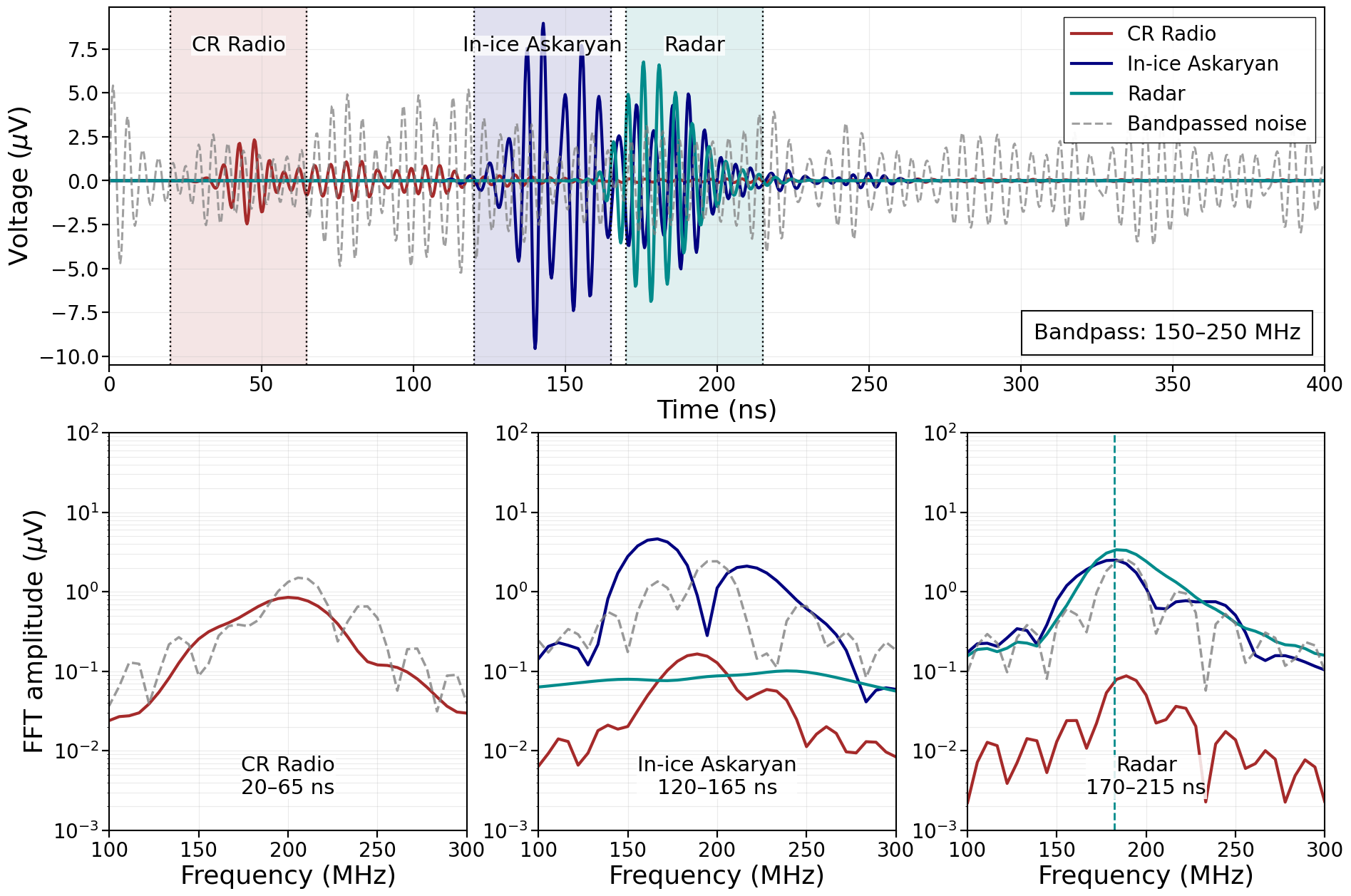}
    \caption{ Top: First order demonstration for voltage estimates at RET-CR within a bandpass of 150--250 MHz, for a cosmic ray primary proton of energy $10^{18}$eV and a vertical shower. Bottom: Signal Spectrum for the corresponding time windows corresponding to the three signal components - CR Radio (left), In-ice Askaryan (center), and radar signal (right).}
    \label{fig:voltage}
\end{figure*}
With RET-CR stations distributed across multiple azimuthal directions, as illustrated in Figure \ref{fig:exp_lay}, different combinations of radio and radar signal strength are expected to be observed at the receivers, which is geometry dependent.  Figure \ref{geoms} presents a heat map of, 
\begin{equation}
\text{Power Ratio} = 10* \log_{10}\left(\frac{P_{\mathrm{radar}}}{P_{\mathrm{Askaryan}}}\right) \text{dB},
\end{equation}
evaluated within the narrow frequency band of 170–200~MHz to identify geometries in which radar echoes are distinguishable from the Askaryan radio emission. Restricting the analysis to a narrow frequency band around the transmitted radar frequency limits the contribution from the in-ice Askaryan emission. 

The map is shown as a function of the shower zenith angle and the azimuthal direction defined with respect to the $+X$ (North) direction. The shower vertex is located at $(10,0,0)$, while the transmitter position is fixed at $(0,0,-10)$. 

We find that, for most geometries of our interest, the radar echo has sufficient power in the vertically polarised channel to exceed the in-ice Askaryan radio component within this narrow frequency band. The radial two-dimensional radio and radar-energy profiles for full bandwidth is depicted in Figure \ref{fig:fluence} in the Appendix \ref{appendix3}.
\subsection{Towards Realistic Signal Modelling}
\label{voltagesec}
A first step towards realistic signal modelling is the conversion of the simulated electric-field signals into voltage traces, which represent the quantity measured by the receiver antennas. This requires an understanding of both the antenna and detector system responses. A complete treatment would further require a detailed frequency-dependent response model and a characterisation of the system noise. Such a comprehensive analysis lies beyond the scope of the present work and is deferred to a dedicated follow-up study.

The primary objective is to predict the signal characteristics in the voltage domain on a first-order basis and assess their detectability against a representative noise level.  For this, a  simplified dipole antenna model is adopted,  described by 
\begin{equation}
L_{\mathrm{eff}}(\Theta,\Phi) = \sin^{2}(\Theta),
\label{eq}
\end{equation}
to approximate the detector response. Here $\Theta$ and $\Phi$ are the polar and azimuthal angles of the incoming signal. The antenna sensitivity is assumed to depend only on the zenith angle. Frequency dependence is not included yet in this simplified model, but we will add an external bandpass filter (100 -- 300 MHz) to match the RET in--ice antennas.  

We also include a thermal noise contribution - the Johnson-Nyquist noise - where the noise amplitudes are drawn from a Rayleigh distribution, consistent with thermal noise in radio-frequency systems; with $V_{\rm rms} = \sqrt{k_B T R \Delta f}$ $\approx$  6.33 $\mu$~V assuming $T = 290\,\mathrm{K}$, RET-CR bandwidth of 200 MHz and resistance of 50 $\Omega$. For a 100 MHz bandpass, the corresponding RMS voltage is $V_{\rm rms}\approx 4.48\,\mu\mathrm{V}$.

To evaluate the antenna response, the signal arrival direction at the detector must first be determined. To obtain the recieving angles incident with our in-ice dipole antennas, raytracing solutions for in-air radio and in-ice Askaryan radio were determined. For in-air radio emission, we calculated the optimal solution from Snell's law, minimized for the least optical path allowed for propagation between the shower maxima in air, as it travels towards the receivers at 10m depths in ice. For the in-ice emission, we propagate the rays between the in-ice shower maxima in ice and the receivers at 10 m depth. In both cases, we use Radiopropa \citep{radiopropa} for raytracing in an in-ice medium with the Greenland firn refractive index profile. Details of the ray-tracing procedure are provided in Appendix \ref{appendix3} in Figure~\ref{fig:ray}. For the radar signal, the ray-tracing implementation within MARES accounts for propagation from each scattering element along the cascade to the receiver antennas \citep{latif}.

A vertical air shower ($\theta = 0^\circ$) with a primary energy of 1 EeV is considered at a distance of 30~m at the location for $r_1$, as shown in Figure~\ref{fig:configuration}(iii). 

The ray-tracing solutions for the in-air emission indicate that the receiving vectors are steeper, corresponding to signals arriving from higher up in the atmosphere. As a result, when the shower maxima occur further up from the ground, the receivers might be less sensitive to the in-air radio signal.

The shower maximum for the in-ice Askaryan emission occurs within the ice. The corresponding receiving vectors at the detector have polar angles that are closer to the antenna's sensitivity, resulting in a stronger response. 

Figure \ref{fig:voltage} illustrates the bandpass-filtered time-domain voltage signals within a bandpass of 150--250 MHz, from the three emission components: the cosmic-ray radio signal, the in-ice Askaryan emission, and the radar echo (at the transmitted power of 544~W),  together with the added thermal noise.  To investigate the frequency characteristics independently, the bandpass-filtered time-domain traces are divided into three windows corresponding to the three signals, and the respective spectra are in the bottom panel. The thermal-noise spectrum extracted from the corresponding time window is shown for comparison.

We observe that in the time domain voltage estimates,  the in-ice Askaryan emission and the radar echo can be visible above the thermal-noise background. The frequency spectra further demonstrate that each time window is dominated by its corresponding emission mechanism, that is the in-ice Askaryan emission dominates the Askaryan window, and the radar echo dominates the radar window around the transmitted frequency, despite the presence of  contributions from the other emission components. This behaviour is geometry-dependent and in this particular geometry, we see that the cosmic-ray radio contribution is comparatively weaker.
\section{Conclusion}
We presented a simulation-based study of the complete signal expected at in-ice receivers for the RET-CR geometry, combining contributions from both in-air and in-ice radio emission together with radar echoes. This was achieved using multiple simulation frameworks: FAERIE (CORSIKA and Geant4) for cosmic-ray air showers and particle energy deposition in ice, and MARES for modelling radar echoes.

We further examined the signal characteristics across different transmitted powers and analysed the full frequency-domain response for various geometrical configurations. We showed that the radar, Askaryan, and in-air radio signals can be distinguished through their frequency content and arrival times, and demonstrated how the geometry of the air shower plays a key role in determining the observed signal.

We find that the radar signal should be detectable within the detector bandwidth, depending on the cascade geometry, at the transmitted powers of 136 W and 544 W used during the 2024 RET-CR campaign. Although the unfiltered Askaryan and cosmic-ray radio signals are substantially stronger than the radar echo, much of their power lies outside the receiver bandwidth, allowing the receiver to pin down the radar signal. 

The search for radar signals from cosmic ray-induced in-ice secondary cascades is currently underway within the collaboration, and this simulation-based analysis, together with the characterisation of the arrival times and frequency content of the different signal components, will guide that search \citep{Frikken2025RETCR}.

\appendix
\section{CORSIKA simulations}
\label{appendix1}
The CORSIKA-77500 with QGSJET-II simulations were performed using realistic geomagnetic field components for Greenland ($B_x = 8.46,\mu\text{T}$ and $B_z = 54.14,\mu\text{T}$) at an observation altitude of 3.2 km. Thinning was applied with parameters $\varepsilon = 10^{-6}$, $w_{\text{max}} = 100$, and $r_{\text{max}} = 0$.
\\
The Greenland ice profiles are taken from previous studies, as described in \citep{article23}. 
The atmospheric model is represented by five consecutive layers. The lower four layers follow an exponential profile,
\[
T(h)=a_i+b_i e^{-h/c_i}, \qquad i=1,2,3,4,
\]
while the uppermost layer is described by
\[
T(h)=a_5-b_5 h/c_5.
\]
The model is derived from GDAS data for Summit Station, as prescribed in \citep{article21}. The corresponding parameters are listed below.
\begin{table}[!htbp]
\centering
\begin{tabular}{|c|c|c|c|c|}
\hline
Layer & ATMLAY & $a_i$ & $b_i$ & $c_i$ \\
\hline
1 & $0.00\times10^{0}$  & $-1.81\times10^{2}$ & $1.22\times10^{3}$ & $9.52\times10^{5}$ \\
2 & $3.53\times10^{5}$  & $-9.28\times10^{1}$ & $1.14\times10^{3}$ & $8.52\times10^{5}$ \\
3 & $8.99\times10^{5}$  & $1.65$              & $1.21\times10^{3}$ & $6.50\times10^{5}$ \\
4 & $2.71\times10^{6}$  & $5.97\times10^{-4}$ & $9.39\times10^{2}$ & $7.08\times10^{5}$ \\
5 & $1.00\times10^{7}$  & $1.13\times10^{-2}$ & $1.0$              & $1.0\times10^{9}$ \\
\hline
\end{tabular}
\caption{Five-layer atmospheric density parameters.}
\end{table}

\section{Additional Figures:}
\begin{figure}
  \centering
  \includegraphics[width=.84\columnwidth]{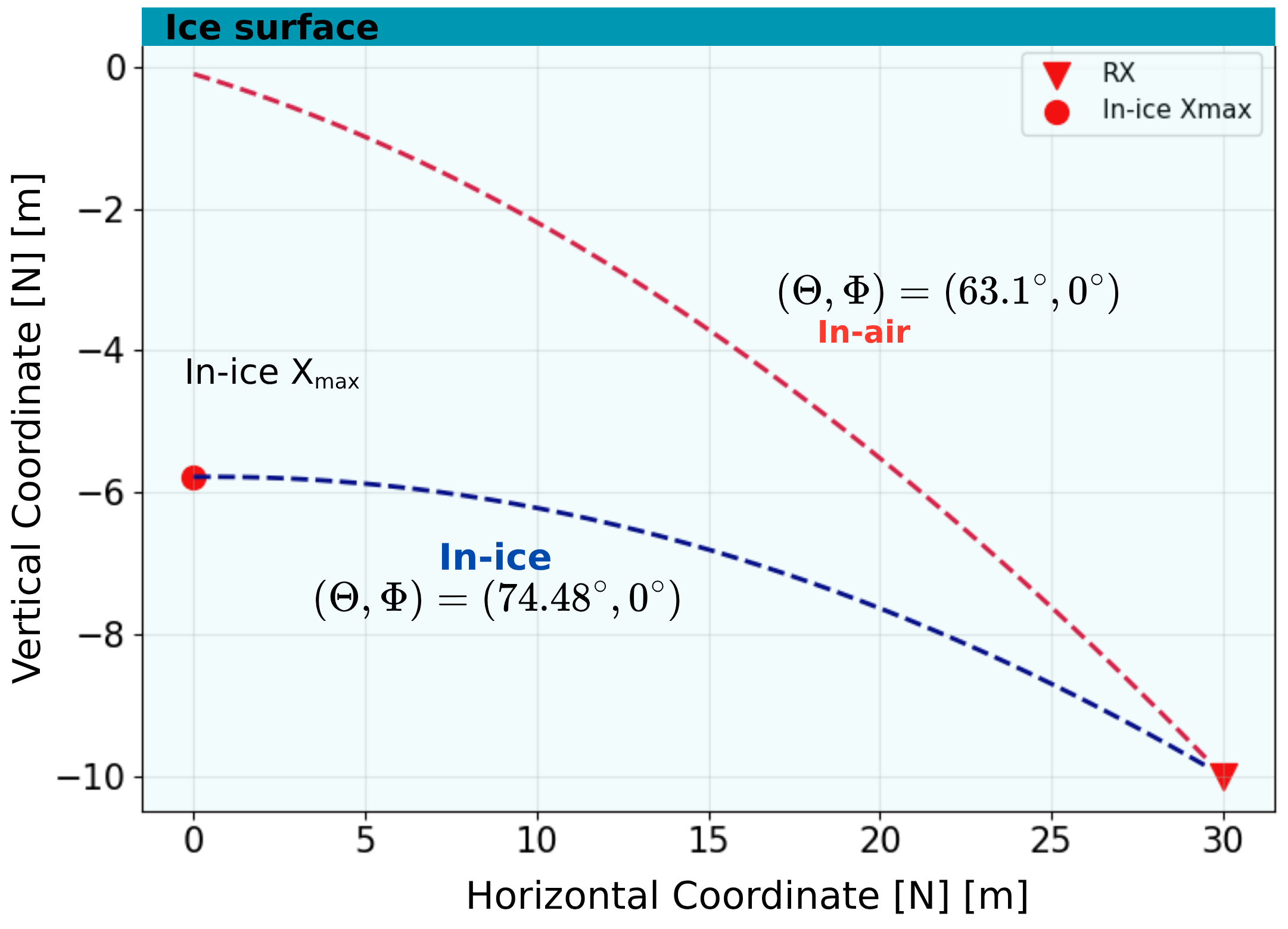}
    \caption{Raytracing solutions for section \ref{voltagesec}}
  \label{fig:ray}
\end{figure}
\begin{figure*}[!htbp]
    \centering
    \includegraphics[width=\textwidth]{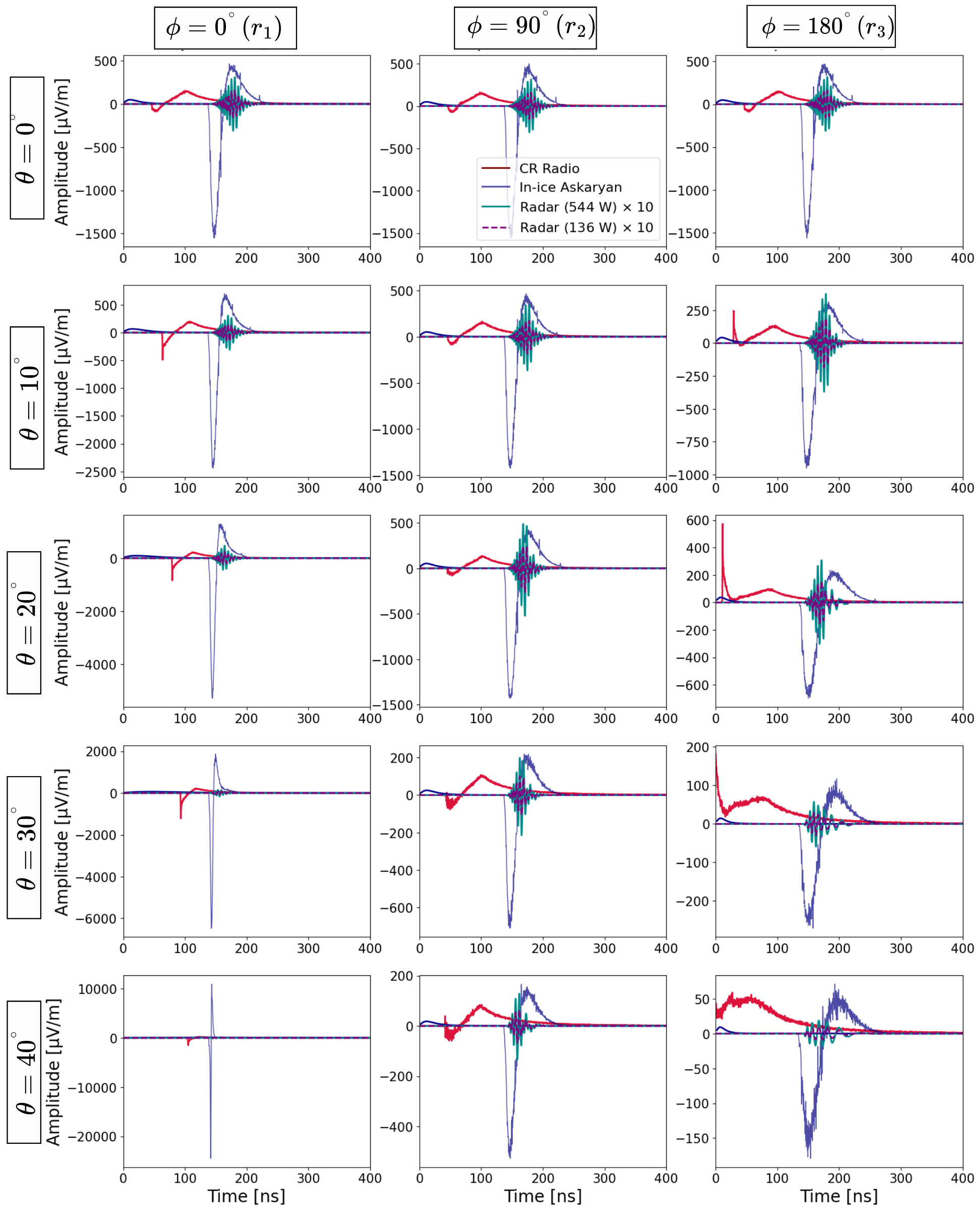}
    \caption{The simulated electric fields at locations corresponding the geometry in Figure \ref{fig:configuration}(iii), and the spectrum for the same is shown in Figure \ref{fig:spec_all}.  All antennas are at a distance of 30 m from the shower vertex at [10, 0, 0] on the surface of ice. The transmitter is positioned at [0, 0,-10]. The energy of the CR primary is $10^{18} eV$ and the radar is powered at 136W and 544 W.  The amplitudes are scaled for clarity of presentation and are not directly comparable in this figure.}
    \label{fig:time_all}
\end{figure*}
\begin{figure*}[!htbp]
    \centering
    \includegraphics[width=\textwidth]{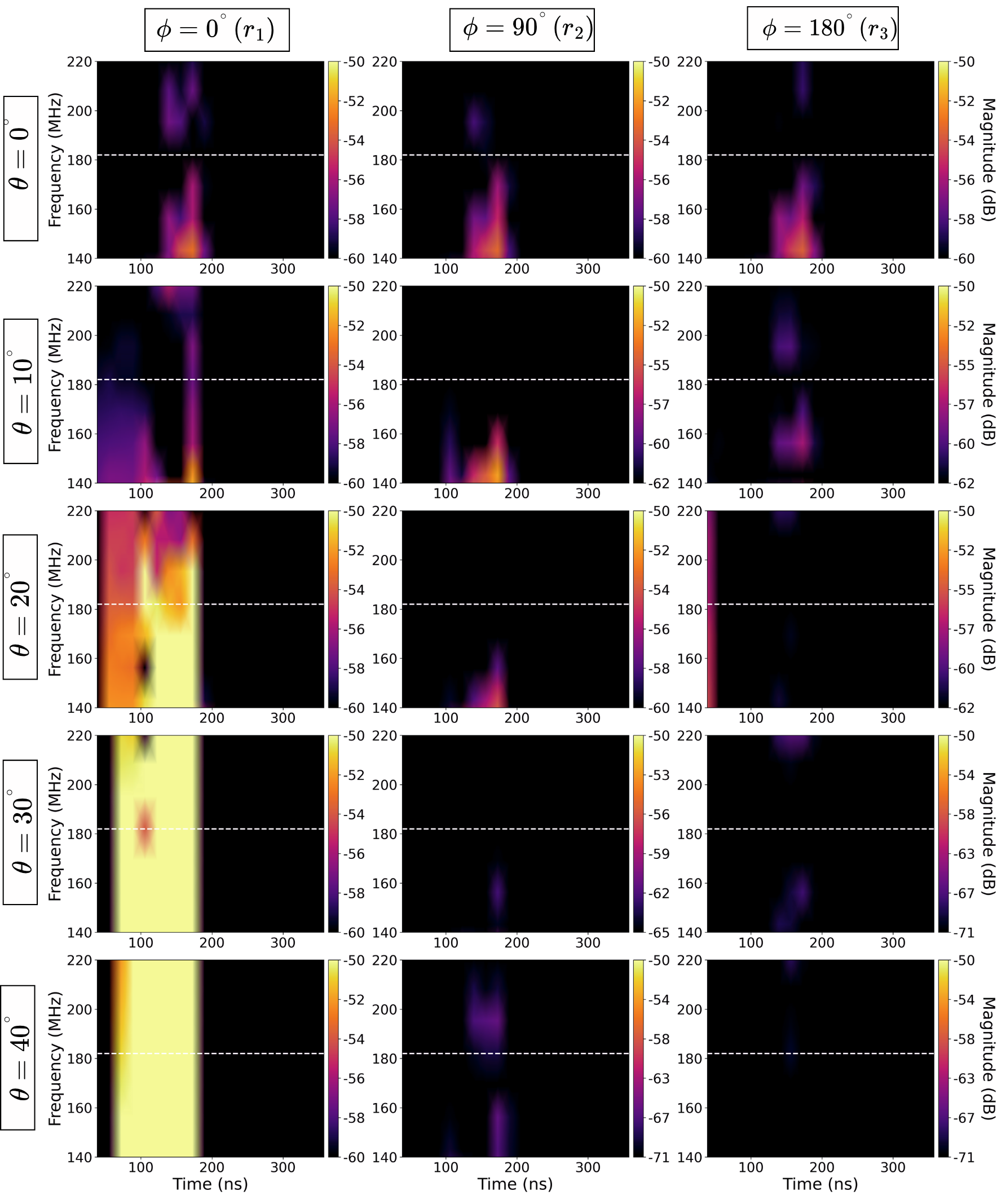}
    \caption{The spectrogram with only in-air and in-ice Askaryan radio contributions, corresponding to three receiver azimuthal directions, $r_1$, $r_2$, and $r_3$, for the simulated signals in Figure \ref{fig:time_all} and spectra in Figure \ref{fig:spec_all}.  The configuration is depicted in Figure  \ref{fig:configuration}(iii). The energy of the CR primary is $10^{18}$ eV and the radar is powered at 544 W.}
    \label{fig:spectrographs_noradar}
\end{figure*}
\begin{figure*}[!htbp]
    \centering
    \includegraphics[width=\textwidth]{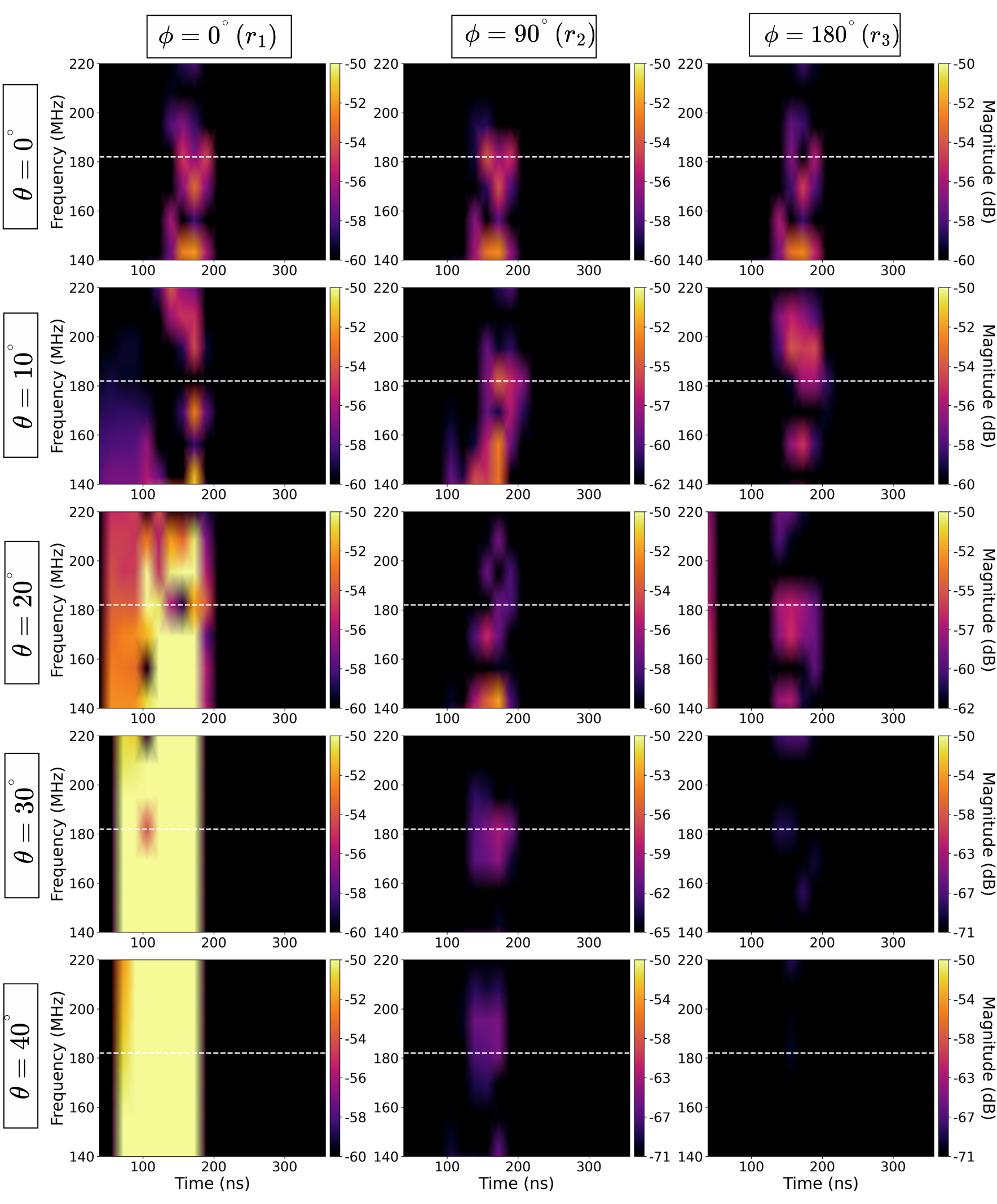}
    \caption{The spectrograms with all radio and radar contributions, corresponding to three receiver azimuthal directions, $r_1$, $r_2$, and $r_3$, for the simulated signals in Figure \ref{fig:time_all} and spectra in Figure \ref{fig:spec_all}.  The configuration is depicted in Figure  \ref{fig:configuration}(iii). The energy of the CR primary is $10^{18}$ eV and the radar is powered at 544 W.}
    \label{fig:spectrographs_radar}
\end{figure*}
\begin{figure*}[!htbp]
    \centering
    \includegraphics[width=0.90\linewidth]{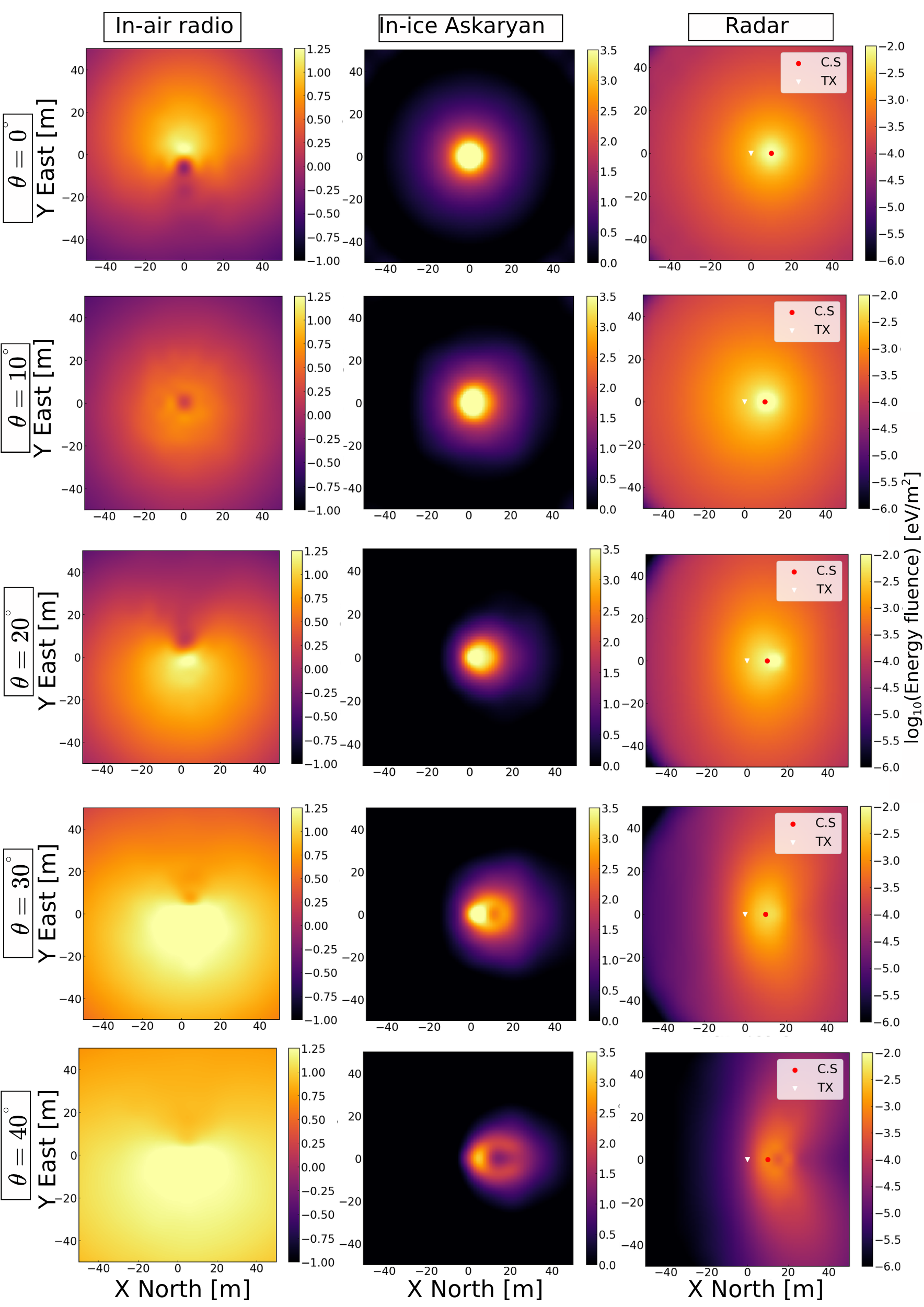}
    \caption{Radial energy footprint from i) in-air radio ii) in-ice Askaryan iii) radar signals \textit{(left to right)} at a depth of 10 m in ice for 100 PeV for shower of zenithal directions $0^{\circ}$,$10^{\circ}$,$20^{\circ}$, $30^{\circ}$, and $40^{\circ}$ \textit{(top to bottom)}}
    \label{fig:fluence}
\end{figure*}

\label{appendix3}
In this section, we present the time-domain signals for the detector geometry shown in Figure~\ref{fig:configuration}(iii) (Figure~\ref{fig:time_all}) and the corresponding frequency spectra (Figure~\ref{fig:spec_all}). The corresponding spectrograms: (i) with only radio signals is in Figure \ref{fig:spectrographs_noradar}, and (ii) with radio and radar echoes together is in Figure \ref{fig:spectrographs_radar}. We also present the two-dimensional radio and radar energy fluence footprints in Figure \ref{fig:fluence}. The ray-tracing procedure used for the voltage estimation is illustrated in Figure~\ref{fig:ray}. 
\bibliographystyle{cas-model2-names}

\bibliography{cas-refs}




\end{document}